\documentclass[twocolumn]{aastex631}

\usepackage{xspace}
\usepackage{enumitem}

\newcommand{\fluxcgs}{erg\,cm$^{-2}$\,s$^{-1}$}

\newcommand{\lat}{{\it Fermi}-LAT\xspace}

\newcommand{\erass}{\texttt{eRASS1}\xspace}
\newcommand{\wisea}{\texttt{WISEA}\xspace}
\newcommand{\erosita}{\textit{eROSITA}\xspace}
\newcommand{\wise}{\textit{WISE}\xspace}
\newcommand{\rosat}{\textit{ROSAT}\xspace}
\newcommand{\sumss}{\texttt{SUMSS}\xspace}
\newcommand{\nvss}{\texttt{NVSS}\xspace}
\newcommand{\first}{\texttt{FIRST}\xspace}

\shorttitle{TeV-emitting blazars candidates from the \erosita survey}
\shortauthors{Metzger, Gokus \& Errando}

\begin{document}

\title{New TeV-emitting BL~Lac candidates from the eROSITA X-ray survey}

\author[0000-0003-3585-3356]{Cassidy Metzger}
\affiliation{Department of Physics \& McDonnell Center for the Space Sciences, Washington University in St. Louis, One Brookings Drive, St. Louis, MO 63130, USA}
\correspondingauthor{C. Metzger, A. Gokus, M. Errando}
\email{c.m.metzger@wustl.edu, dr.andrea.gokus@gmail.com, errando@wustl.edu}
\author[0000-0002-5726-5216]{Andrea Gokus}
\affiliation{Department of Physics \& McDonnell Center for the Space Sciences, Washington University in St. Louis, One Brookings Drive, St. Louis, MO 63130, USA}
\author[0000-0002-1853-863X]{Manel Errando}
\affiliation{Department of Physics \& McDonnell Center for the Space Sciences, Washington University in St. Louis, One Brookings Drive, St. Louis, MO 63130, USA}

\begin{abstract}

TeV-emitting BL Lac type blazars represent the extreme end of the blazar population. They are characterized by relatively weak jets and radiatively inefficient accretion disks. Particles accelerated in these jets experience fewer radiative losses, allowing them to reach energies beyond the TeV scale and produce TeV gamma-ray emission. The study of TeV blazars is constrained by the limited number of known sources in this category. Currently, only 56 high synchrotron-peaked BL Lacs have been detected at energies above 0.1\,TeV.
Searches for TeV emission from BL~Lacs typically target sources with bright X-ray emission and a synchrotron peak at or above 1\,keV. The recently released \erass catalog by the \erosita collaboration, which covers half of the sky, represents the deepest X-ray survey in the soft X-ray band to date. Utilizing the \erosita survey, combined with infrared data from \wise and archival radio observations, we have identified 121 TeV-emitting blazar candidates. Our search introduces selection criteria based on the radio to infrared that remove quasar-like objects that have similar infrared spectra and X-ray fluxes as TeV-emitting BL~Lacs. In our search, we find 23 objects that had not been detected in the \rosat X-ray survey and 11 that have not been previously associated with blazars. 
The candidates resulting from our search are suitable for follow-up observations with currently operating imaging atmospheric Cherenkov telescopes, as well as future facilities like the CTAO Observatory.
\end{abstract}


\section{Introduction} \label{sec:intro}
Active galactic nuclei (AGN) are observed to launch relativistic jets powered by accretion onto a central supermassive black hole
\citep[e.g., see the recent review by][]{Blandford_2019}. These collimated streams of matter and radiation occur in approximately 10\% of AGN \citep{Kellermann1989, Urry, Ivezic2002}. The prevailing understanding of the physical structure of an AGN is that they consist of a central supermassive black hole ($>10^6 M_\odot$) with a surrounding accretion disk that emits ultraviolet and soft X-ray radiation. A warped ring of gas and dust, often depicted as a torus, encircles the accretion disk and obscures UV and optical radiation when viewed edge-on. Relativistic jets are launched perpendicular to the plane of the accretion disk and can be seen across all wavelengths. A unified view of AGN proposed by \citet{Urry} posits that the various classes of AGN actually represent different perspectives of the same underlying object, depending on the observer's line of sight. Blazars exhibit the most extreme properties of all AGN. They are objects that we observe from a very small viewing angle, that is, with their jets pointed toward us, and exhibit large-amplitude flux variability across the electromagnetic spectrum. Their spectral-energy distribution (SED) typically has two distinct parts: a broad synchrotron component at lower energies and a second high-energy component that can extend into the gamma-ray band. In flat-spectrum radio quasars (FSRQs) and low-frequency peaked BL~Lac-type blazars, the synchrotron peak typically appears in the infrared to UV wavelengths. However, in high-frequency peaked BL~Lac-type blazars (HBLs), the synchrotron peak is located in the X-ray region ($\nu_\mathrm{peak} \geq 10^{15}$) \citep{PadovaniGiommi}. The second component is attributed to inverse Compton scattering, and in HBLs it can extend from X-rays to TeV gamma rays. 
The high-energy component of the SED of some HBLs peaks 
beyond 10\,TeV. These objects are often labeled extreme high-frequency peaked BL~Lacs \citep[eHBLs,][]{2001A&A...371..512C,Bonnoli_2015, Arsioli_2020, arsioli_2025}, and their emission challenges 
standard synchrotron self-Compton models for BL Lacs \citep{2018MNRAS.477.4257C}. These BL~Lacs with radiative output peaking in the TeV band have interesting implications for the study of jet phenomenology, particle acceleration, and production of ultra-high-energy cosmic rays and neutrinos, and can be used as probes of extragalactic background light \citep{2006Natur.440.1018A,2007A&A...471..439M,2015ApJ...812...60B, IceCube_2018, Padovani_2018, Chang_2022} and cosmic magnetic fields \citep{2010Sci...328...73N,2015PhRvL.115u1103C,2015ApJ...814...20F, 2017ApJ...835..288A,2023A&A...670A.145A,2023ApJ...950L..16A}. 
However, due to the limited sensitivity and survey capabilities of TeV observatories \citep{2008RPPh...71i6901A,2022Galax..10...21S,2023hxga.book..144E}, the number of eHBLs currently identified and characterized ranges from one to two dozen \citep{2020NatAs...4..124B, Nievas_Rosillo_2022}.

{The population of known gamma-ray-emitting blazars remains notably incomplete. 
The \lat all-sky blazar survey is only complete at energy fluxes $>3\times 10^{-12}$\,\fluxcgs\ \citep{2020ApJ...896....6M}, despite the LAT being able to significantly detect sources an order of magnitude fainter. 
Traditionally, blazars are identified as compact flat-spectrum radio sources with a continuum-dominated optical counterpart. Other characteristics, such as a point-like X-ray counterpart, detection of gamma-ray emission, optical polarization, or flux variability across radio, optical, or X-ray bands, further substantiate a blazar identification.
Searches for new TeV-emitting blazars typically follow TeV flux predictions based on radio, optical and X-ray surveys \citep[e.g.,][]{2002A&A...384...56C,2005A&A...434..385G, Zhu_2021}. However, these searches are often limited by incomplete coverage of surveys at these wavelengths or by source confusion. Examples of these limitations are the growing number of blazars that have been serendipitously detected at low galactic latitudes in the GeV band \citep[e.g.,][]{2000ApJ...542..740M,2010ApJ...718L.166V,2012ApJ...746..159K} and at TeV energies
\citep[e.g.,][]{2011A&A...529A..49H,2013ApJ...776...69A}. 
These blazars do not appear in conventional blazar catalogs because optical and X-ray point source catalogs tend to be incomplete at low Galactic latitudes or exclude the Galactic plane altogether \citep[e.g.,][]{2008ApJS..175...97H,Massaro2009}.
Recent efforts have aimed to identify new TeV-peaked blazar candidates using deeper and more complete radio, infrared, optical, and X-ray catalogs \citep[e.g.,][]{Massaro_2013,Costamante_2020, Arsioli_2015, Chang_2017} with the goal of observing and potentially detecting such candidates with ground-based gamma-ray observatories.}

In this paper, we use the \erass X-ray survey of the \erosita mission \citep{Merloni_2024} to identify new TeV-emitting BL Lac candidates. The \erass\ catalog covers the western galactic hemisphere (half of the sky) in the 0.2--2.3\,keV energy range with a deeper sensitivity than any previous wide X-ray survey, resulting in a source density of $\sim 45\,\textrm{deg}^{-2}$, more than an order of magnitude higher than the \rosat \texttt{2RXS} catalog \citep{2016A&A...588A.103B}.

To identify new candidate TeV-emitting BL~Lacs we select infrared sources in the {\texttt{AllWISE} (\wisea) catalog \citep{cutri_2013} which combines data from the \wise \citep{Wright_2010} and \textit{NEOWISE} \citep{Mainzer_2011} surveys.} We require that sources occupy the region in the \wise color-color diagram where gamma-ray blazars are found \citep{Massaro_2012, Arsioli_2015, Chang_2017} and cross-match them with their X-ray counterparts in the \erosita \erass survey to select objects with a luminous synchrotron component peaking in the X-ray band. 

\section{TeV-emitting BL~Lac-type blazars}\label{sec:mesh:2}

The first step in our search for new candidate TeV-emitting HBLs is to establish the infrared and X-ray properties of known examples of this source class. BL~Lac-type blazars are the most numerous class of extragalactic sources detected at TeV energies, with 68 currently detected \citep[\texttt{TeVCAT}\footnote{\url{http://tevcat2.uchicago.edu}},][]{2008ICRC....3.1341W}. A significant portion of these, totaling 56, are classified as HBLs.
The localization uncertainty for extragalactic TeV sources is typically $\gtrsim 10''$ \citep[e.g.,][]{2013ApJ...776...69A}. To determine the infrared and X-ray properties of known TeV-emitting HBLs, we searched for their optical counterparts in the \texttt{Simbad}\footnote{\url{https://simbad.cds.unistra.fr/simbad}} database to obtain better localizations. Using these optical coordinates, all 56 TeV HBLs were found to have an infrared counterpart in the \wisea catalog within $5''$.

\begin{figure}
    \centering
    \includegraphics[width=0.5\textwidth]{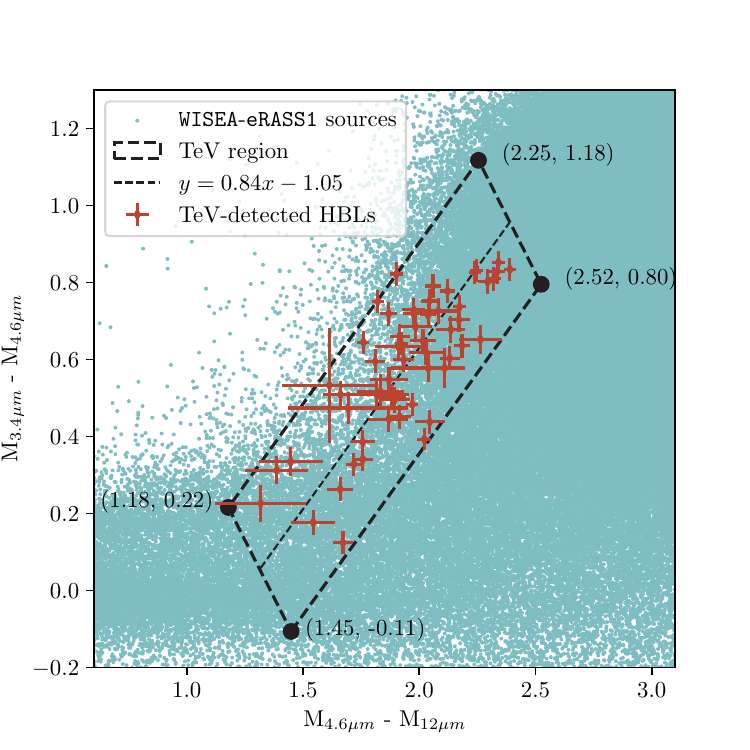}
    \caption{The \wise $3.4\text{--}4.6\text{--}12\, \mu \text{m}$ color-color plot for the 56 TBLs selected (red crosses). The dashed line corresponds to the regression line described in Section~\ref{sec:mesh:2}. The black dashed rectangle corresponds to the maximum allowed separation (0.21) of the known TeV HBLs from this line. The turquoise points correspond to all \wisea sources found within the \erosita footprint.}
    \label{fig:cc_diagram}
\end{figure}

The first criterion we will use to identify TeV HBL candidates is based on their infrared colors. \citet{2011ApJ...740L..48M} showed that blazars occupy a strip in the \wise $3.4-4.6-12\,\mu$m color-color diagram, making them distinguishable from other extragalactic sources. \citet{Massaro_2012} demonstrated that gamma-ray emitting BL~Lacs can further be separated from flat spectrum radio quasars based on their infrared colors. The infrared colors of the \wise counterparts for the 56 known TeV HBLs are shown inFigure~\ref{fig:cc_diagram}).  
We find that TeV HBLs cluster along the $y = 0.84x -1.05$ line in the \wise $3.4-4.6-12\,\mu$m color-color diagram. 
The farthest object from the best-fit line is at a distance of 0.21. In our search, we will require that new TeV HBL candidates be within a distance of less than 0.21 from the best-fit line determined from known TeV HBLs. The rectangle with boundaries shown in Figure \ref{fig:cc_diagram} defines the region where TeV HBLs are found. In line with the findings described in \citet{Arsioli_2015} and \citet{Chang_2017}, our search region in infrared color space is wider compared to that in \citet{Massaro_2013} to produce a more complete HSP sample. 

\begin{figure}
    \centering
    \includegraphics[width=0.5\textwidth]{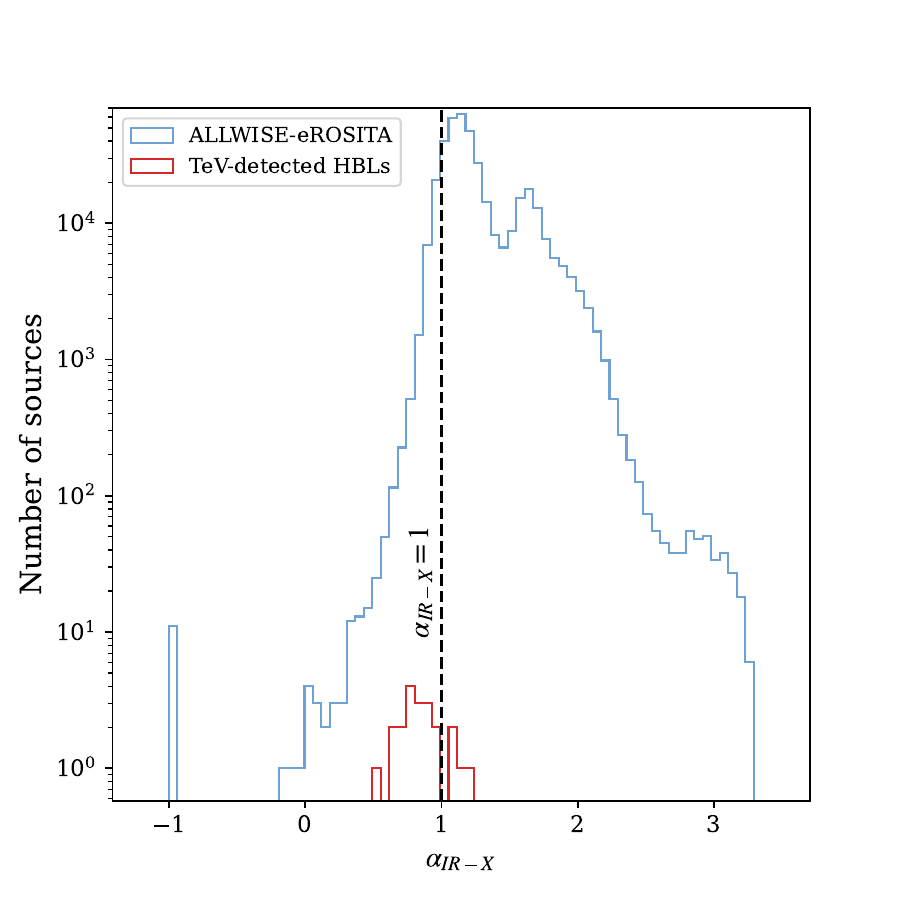}
    \caption{The distribution of $\alpha_{\text{IR-X}}$ for all sources found in \erass and \wisea (blue) and the 21 TBL sources found in the \erass footprint (red). The dashed black line represents the our selection criterion of $\alpha_{\text{IR-X}} \leq 1$.}
    \label{fig:alpha_ir-x}
\end{figure}

The next criterion for selecting TeV HBL candidates involves assessing their effective spectral index (or spectral slope) between the infrared and X-ray frequencies. First, we use the \wise coordinates of known TeV HBLs to find their X-ray counterparts in the \erosita \erass catalog. We match infrared sources with their closest X-ray counterpart as long as the separation is less than $5''$, which is the average positional error in the \erass catalog \citep{Merloni_2024}, considering both statistical and systematic uncertainties. If multiple counterparts are found, only the closest spatial match is considered. Only 21 of the 56 known TeV HBLs are located in the western galactic hemisphere covered in the public \erass data. 
To distinguish between HBLs and LBLs, we define the infrared to X-ray effective spectral index:
\begin{equation}\label{eq:alpha_irx}
    \alpha_{\text{IR-X}} = - \frac{\log(F_{\text{3.4$\mu$m}}/F_{\text{$0.2-2.3$\,keV}})}{\log(\nu_{\text{3.4$\mu$m}}/\nu_{\text{$1$\,keV}})}
\end{equation}
The distribution of $\alpha_{\text{IR-X}}$ for all \wise sources with an \erosita counterpart is shown in Figure~\ref{fig:alpha_ir-x}. 
We require that new TeV HBL candidates have $\alpha_{\text{IR-X}}\geq 0$, as that cut maximizes the number of retained TeV HBLs while minimizing the number of potential candidates to be considered. 
{We note that four known TeV HBLs in the \erass footprint have $\alpha_{\text{IR-X}} < 0$: PKS~1440-389, 1ES~1215+303, TXS~1515-273, and PKS~0301-243.} 
However, the $\alpha_{\text{IR-X}}$ distribution for a general population of sources peaks at negative values, and relaxing the infrared to X-ray spectral index cut to include these four sources would significantly increase the number of seed infrared sources in our search and result in a higher potential for false positives. The observational properties of known TeV-emitting HBLs are shown in Table~\ref{table:TeV}.

\begin{figure*}
    \centering
    \includegraphics[width=1.0\textwidth]{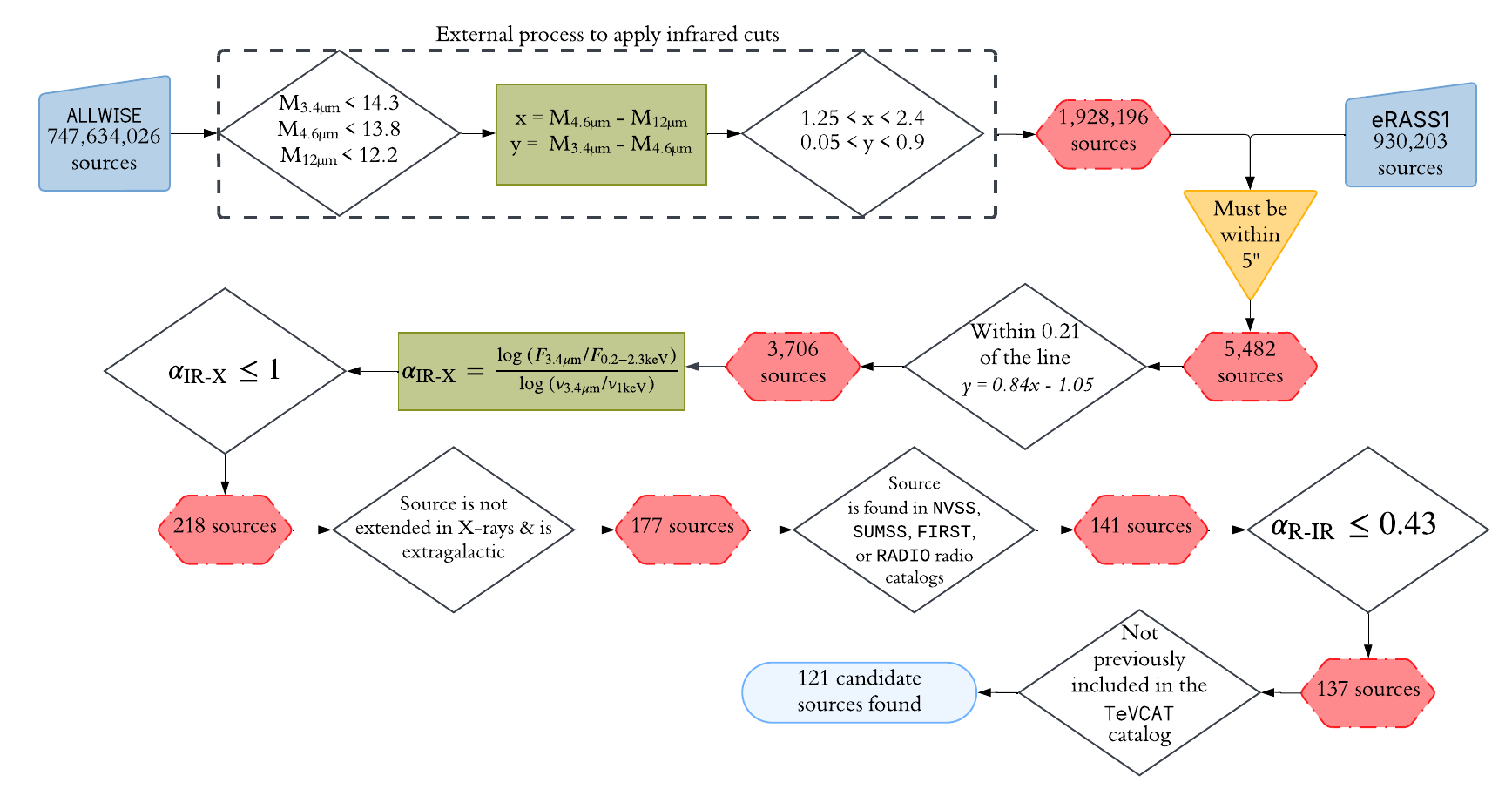}
    \caption{Description of the algortihm followed to identify TeV-emitting HBL candidates. The input surveys \wisea and \erass are shown at the top in light blue. The resulting outcome of 135 objects is shown at the bottom, also in light blue. The number of objects that make it through each stage of filtering is shown in red boxes. Green boxes indicate parameters that are calculated based on catalog inputs.}
    \label{fig:algorithm}
\end{figure*}

\section{Search for new TeV HBL candidates}\label{sec:2}
\begin{figure}[htb]
    \centering
    \includegraphics[width=0.5\textwidth]{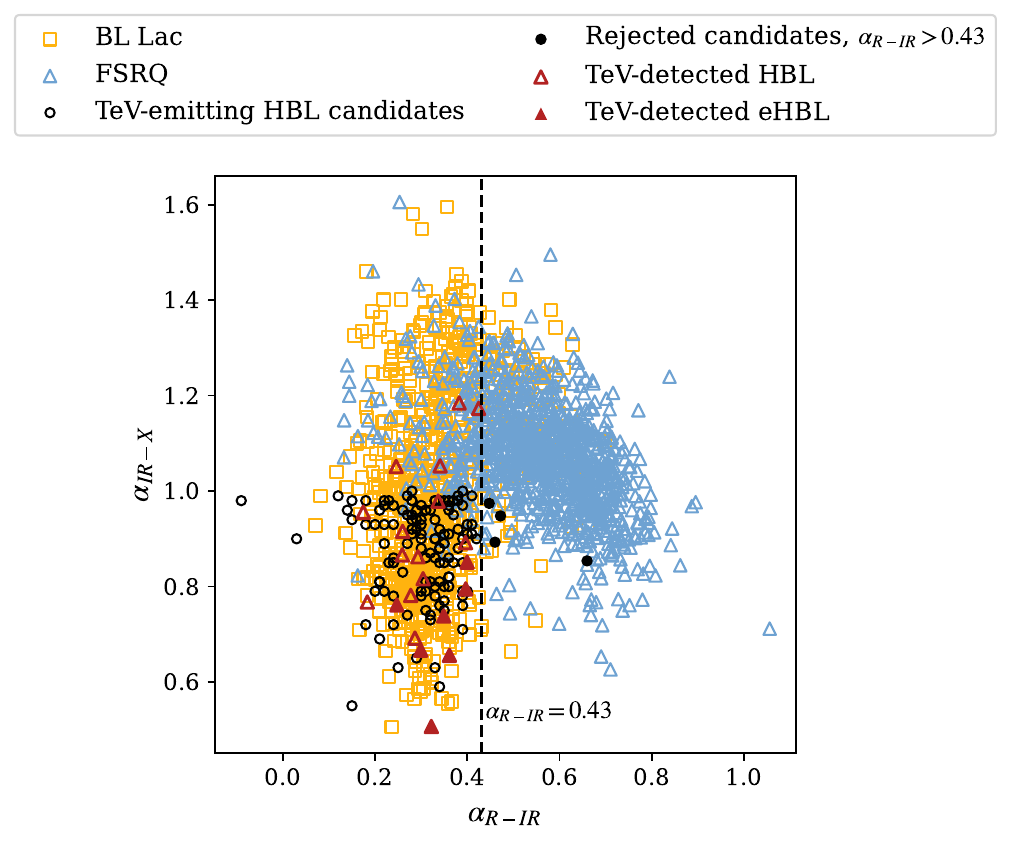}
    \caption{The ratio between the radio and IR fluxes, normalized by the ratios of their frequencies, plotted against $\alpha_{\text{IR-X}}$ defined in Equation~\ref{eq:alpha_irx}. BL~Lac objects in ROMA-BZCAT are displayed as yellow triangles, while FSRQs are displayed as blue triangles. The 21 TeV-detected HBLs within the \erosita footprint are shown as unfilled red triangles, while the ones designated as extreme HBLs (eHBLs) in \texttt{TeVCat} are shown as filled red triangles. The 135 candidate sources identified in our study are shown as open black circles. Sources that are have passed all previous candidacy criteria but have $\alpha_{\text{R-IR}} \leq 0.43$ are designated as green circles outlined in black.}
    \label{fig:sourcetypes}
\end{figure}

Based on the infrared and X-ray properties of known TeV-emitting HBLs, we define the following procedure to identify new candidate TeV HBLs (Figure~\ref{fig:algorithm}). Our candidate list includes all sources that meet the following criteria:

\begin{enumerate}[leftmargin=*,itemsep=0pt,topsep=0pt,label=\Roman*.] 
\item Infrared \wisea source detected with Vega magnitudes of less than 14.3, 13.8, and 12.2 at $3.4 \mu \text{m}$, $4.6 \mu \text{m}$, and $12\mu \text{m}$, respectively; 
\item \wisea source has infrared colors similar to known gamma-ray blazars:  $1.25\,\text{mag}< M_{4.6\mu \text{m}} - M_{12\mu\text{m}} < 2.4\,\text{mag}$, $0.05\,\text{mag} < M_{3.4 \mu \text{m}} - M_{4.6 \mu~\text{m}} < 0.9\,\text{mag}$; 
\item \wisea source falls within a distance of 0.21 from the best-fit line in the infrared color-color diagram where known TeV HBLs are located (Figure~\ref{fig:cc_diagram}), as defined in Section~\ref{sec:mesh:2};
\item Infrared \wisea source has an X-ray counterpart in the \erass catalog within $5''$; 
\item Source has an effective infrared to X-ray spectral index $\alpha_{\text{IR-X}} \leq 1.0$ (Figure~\ref{fig:alpha_ir-x} and Equation~\ref{eq:alpha_irx}); 
  \item X-ray source is not listed as extended in \erass,  guaranteeing that the X-ray object is compatible with a point source as expected for blazars;
  \item Source has a galactic latitude $|b| \geq 10^\circ$ to avoid the larger level source contamination expected near the galactic plane. 
  \item Source has a counterpart in {the \texttt{NVSS} \citep{Condon_1998}, \texttt{SUMSS} \citep{Mauch_2013}, \texttt{FIRST} \citep{Becker_1995} or \texttt{RADIO}\footnote{\url{https://heasarc.gsfc.nasa.gov/w3browse/master-catalog/radio.html}} radio catalogs}. 
  \item The radio-to-infrared effective spectral index $\alpha_{\text{R-IR}}$ of the source is {smaller than 0.43} (Figure~\ref{fig:sourcetypes} and Equation~\ref{eq:alpha_rir}). 
\end{enumerate}

The criteria and procedures for identifying TeV-peaked BL Lac candidates are summarized in Figure \ref{fig:algorithm}. Criteria I-III focus on selecting sources with infrared characteristics similar to those of known TeV-emitting HBLs. 
 
All 56 known TeV-emitting HBLs have WISE magnitudes brighter than 14.3\,mag, 13.8\,mag, and 12.2\,mag in the 3.4\,$\mu$m, 4.6\,$\mu$m, and 12\,$\mu$m bands, respectively (I).
The slope of the jet-dominated synchrotron emission in the infrared band is effectively captured by infrared colors, with candidates required to have colors within these ranges (II): $1.25 < M_{4.6\mu \text{m}} - M_{12\mu\text{m}} < 2.4$ and $0.05 < M_{3.4 \mu \text{m}} - M_{4.6 \mu\text{m}} < 0.9$. The color-color criterion is refined by ensuring that candidate sources fall within 0.21 distance from the best-fit line in the infrared color-color diagram where the known TeV HBLs are located (III, Figure \ref{fig:cc_diagram}). The objects that pass these filters are all significantly detected by \wise\ with infrared colors resembling known TeV-emitting HBLs.

These candidates are then matched with their closest X-ray counterparts in the \erass catalog (IV). We consider X-ray counterparts as long as the separation between the \wisea coordinates and the \erosita X-ray counterpart is smaller than the average \erass positional error of $5''$. Next, sources with $\alpha_{\text{IR-X}} \geq 0$ are selected to identify objects peaking in the X-ray band (V). X-ray emission from blazars is unresolved at the $18''$ spatial resolution characteristic of \erosita. Thus, we reject objects with \texttt{EXT} parameter $> 0$ (VI), denoting extended X-ray sources. Sources excluded by this criterion are listed in Table~\ref{table:extended}. Objects with galactic latitude $|b| \leq 10^\circ$ are rejected to avoid source confusion (VII).
The presence of a radio counterpart in the \texttt{NVSS}, \texttt{FIRST}, \texttt{SUMSS}, or other \texttt{RADIO} catalogs within $5''$ of the \wise\ infrared source is also required (VIII).  Objects that were excluded due to the lack of a known radio counterpart are listed in Table~\ref{table:noRadio}. The radio flux is used to calculate a radio-to-infrared spectral index: 
\begin{equation}\label{eq:alpha_rir}
    \alpha_{\text{R-IR}} = - \frac{\log(F_{\text{R}}/F_{3.4\mu\text{m}})}{\log(\nu_{\text{R}}/\nu_{3.4\mu\text{m}})}
\end{equation}
where the radio flux is preferentially measured at $\nu_{\text{R}} = 1.4$\,GHz. As shown in Figure~\ref{fig:sourcetypes}, $\alpha_{\text{IR-X}}$ and $\alpha_{\text{R-IR}}$ effectively characterize the synchrotron emission component of blazar candidates, distinguishing FSRQs with synchrotron peaks in the infrared band from BL~Lacs with peaks at higher frequencies. TeV-emitting HBLs display harder effective spectral indices in both parameters, clustering in the top right quadrant of the plot. A final selection cut requires TeV-emitting candidates to have $\alpha_{\text{R-IR}} \leq 0.43$ to exclude candidates whose synchrotron SED resembles those of FSRQs more than BL~Lacs.
The python code with the full source selection algorithm is {publicly available}. \footnote{\url{https://github.com/cassiemetzger/THC-CAT/}}

\begin{deluxetable*}{cccccccccccccccccccccccc} 
\rotate

\tabletypesize{\scriptsize} 

\tablecaption{List of TeV-emitting HBL candidates identified in our study, ordered by RA. The complete list of 135 objects can be accessed online in machine-readable format.}

\tablehead{\colhead{THC} & \colhead{WISEA} & \colhead{1eRASS} & \colhead{$F_{3.4\mu \text{m}}$} & \colhead{$F_{0.2-2.3 \text{keV}}$} & \colhead{$F_{2-5 \text{keV}}$} & \colhead{$F_R$} & \colhead{$R_\mathrm{mag}$} & \colhead{$\alpha_{\text{R-IR}}$} & \colhead{$\alpha_{\text{IR-X}}$} & \colhead{\rotatebox{270}{2FHL}} & \colhead{\rotatebox{270}{3FHL}}  & \colhead{\rotatebox{270}{4FGL}} & \colhead{\rotatebox{270}{4LAC-DR3}} & \colhead{\rotatebox{270}{1CGH}} & \colhead{\rotatebox{270}{BZCAT}} & \colhead{\rotatebox{270}{3HSP}} & \colhead{\rotatebox{270}{2WHSP}} & \colhead{\rotatebox{270}{C20}} & \colhead{\rotatebox{270}{D19}} & \colhead{\rotatebox{270}{L25}} & \colhead{\rotatebox{270}{MAR25}} & \colhead{\rotatebox{270}{MAS13}}
\\
\colhead{} & \colhead{} & \colhead{} & \colhead{mJy} & \colhead{$\times 10^{-12} \text{cgs}$} & \colhead{$\times 10^{-12} \text{cgs}$} & \colhead{mJy} 
}

\startdata
J0001-419 & J000132.74-415525.2 & J000132.7-415526 & 0.79 $\pm$ 0.02 & 6.24 $\pm$ 0.25 & 1.89 $\pm$ 0.43 & 13.20  & 17.93 & 0.24 & 0.72 & x & x & x & x & x & x & x & x &  & x & x &  &  \\ 
J0026-460 & J002635.63-460109.9 & J002635.5-460110 & 0.77 $\pm$ 0.02 & 1.32 $\pm$ 0.11 & 0.00 $\pm$ 0.06 & 38.70  & 17.24 & 0.34 & 0.92 &  &  &  &  &  &  &  &  &  &  &  &  &  \\ 
J0041-470 & J004147.02-470136.9 & J004146.9-470136 & 1.15 $\pm$ 0.03 & 3.25 $\pm$ 0.17 & 1.65 $\pm$ 0.38 & 16.60  & 16.81 & 0.23 & 0.85 &  & x & x & x & x &  & x & x &  & x &  &  &  \\ 
J0051-627 & J005116.64-624204.3 & J005116.4-624203 & 1.68 $\pm$ 0.04 & 3.14 $\pm$ 0.14 & 0.99 $\pm$ 0.24 & 43.20  & 15.94 & 0.28 & 0.91 &  &  & x & x & x &  & x & x &  &  &  &  &  \\ 
J0109-403 & J010956.58-402050.9 & J010956.4-402051 & 0.85 $\pm$ 0.02 & 1.56 $\pm$ 0.11 & 0.36 $\pm$ 0.16 & 96.00  & 17.47 & 0.41 & 0.91 & x & x & x & x & x & x & x & x &  & x &  &  &  \\ 
J0116-281 & J011637.06-281146.9 & J011637.1-281145 & 0.75 $\pm$ 0.02 & 0.67 $\pm$ 0.07 & 0.26 $\pm$ 0.14 & 53.60  & 17.99 & 0.39 & 1.00 & x & x & x & x & x & x & x & x &  & x &  &  &  \\ 
J0117-549 & J011751.34-545519.6 & J011751.5-545516 & 0.96 $\pm$ 0.02 & 1.18 $\pm$ 0.09 & 0.24 $\pm$ 0.13 & 7.04  & 17.36 & 0.31 & 0.96 & x &  & x & x & x & x & x & x &  & x &  &  &  \\ 
J0123-231 & J012338.35-231058.7 & J012338.1-231059 & 0.73 $\pm$ 0.02 & 11.92 $\pm$ 0.30 & 4.83 $\pm$ 0.57 & 27.20  & 17.58 & 0.33 & 0.63 & x & x & x & x & x & x & x & x &  & x &  &  &  \\ 
J0143-587 & J014347.41-584551.4 & J014347.6-584549 & 1.30 $\pm$ 0.03 & 8.99 $\pm$ 0.24 & 2.14 $\pm$ 0.35 & 28.30  & 16.19 & 0.27 & 0.74 &  & x & x & x & x & x & x & x &  & x &  &  &  \\ 
J0150-548 & J015044.53-545005.0 & J015044.4-545004 & 0.69 $\pm$ 0.02 & 0.83 $\pm$ 0.08 & 0.13 $\pm$ 0.10 & 43.80  & 17.36 & 0.36 & 0.96 & x & x & x & x & x & x & x & x & x & x &  &  &  \\ 
J0156-530 & J015658.00-530200.0 & J015657.9-530200 & 0.85 $\pm$ 0.02 & 19.69 $\pm$ 0.38 & 10.01 $\pm$ 0.74 & 43.40  & 17.49 & 0.34 & 0.59 & x & x & x & x & x &  & x & x &  & x &  &  &  \\ 
J0209-524 & J020921.62-522922.8 & J020921.6-522922 & 2.54 $\pm$ 0.05 & 12.94 $\pm$ 0.29 & 3.45 $\pm$ 0.44 & 78.70  & 15.77 & 0.30 & 0.78 &  & x & x & x & x &  & x & x &  &  &  &  &  \\ 
J0219-174 & J021905.50-172513.0 & J021905.5-172514 & 1.19 $\pm$ 0.03 & 2.14 $\pm$ 0.12 & 0.92 $\pm$ 0.23 & 62.20  & 16.62 & 0.36 & 0.91 &  & x & x & x & x &  & x &  &  & x &  &  &  \\ 
J0223-112 & J022314.25-111738.4 & J022314.3-111737 & 0.74 $\pm$ 0.02 & 1.61 $\pm$ 0.10 & 0.35 $\pm$ 0.16 & 14.00  & 17.20 & 0.27 & 0.89 &  & x & x & x & x & x & x & x &  &  & x &  &  \\ 
J0244-583 & J024440.30-581954.5 & J024440.2-581955 & 1.67 $\pm$ 0.04 & 10.64 $\pm$ 0.22 & 3.88 $\pm$ 0.39 & 92.60  & 16.07 & 0.35 & 0.75 &  &  &  &  &  &  &  &  &  &  &  &  &  \\ 
J0310-502 & J031034.72-501631.1 & J031034.8-501631 & 0.96 $\pm$ 0.02 & 2.08 $\pm$ 0.09 & 0.63 $\pm$ 0.16 & 12.60  & 17.28 & 0.22 & 0.89 & x & x & x & x & x &  & x & x &  &  &  &  &  \\ 
J0312-223 & J031235.70-222117.2 & J031235.7-222117 & 0.61 $\pm$ 0.01 & 2.17 $\pm$ 0.10 & 0.13 $\pm$ 0.08 & 31.20  & 17.77 & 0.36 & 0.82 &  & x & x & x & x & x & x & x &  &  &  &  &  \\ 
J0325-565 & J032523.53-563544.6 & J032523.2-563544 & 4.39 $\pm$ 0.09 & 22.87 $\pm$ 0.54 & 9.67 $\pm$ 0.55 & 73.30  & 15.37 & 0.24 & 0.78 &  &  &  &  &  &  & x & x &  &  &  &  &  \\ 
J0336-037 & J033623.79-034738.6 & J033623.8-034739 & 1.19 $\pm$ 0.03 & 7.04 $\pm$ 0.20 & 4.57 $\pm$ 0.49 & 53.80  & 17.02 & 0.34 & 0.76 &  & x & x & x & x & x & x & x &  & x &  &  &  \\ 
J0338-287 & J033859.60-284619.9 & J033859.4-284619 & 0.99 $\pm$ 0.02 & 2.24 $\pm$ 0.10 & 0.86 $\pm$ 0.18 & 40.70  & 17.47 & 0.34 & 0.88 &  &  & x & x & x & x & x & x & x &  & x &  &  \\ 
J0352-685 & J035257.48-683117.0 & J035257.5-683118 & 4.44 $\pm$ 0.09 & 8.84 $\pm$ 0.19 & 3.28 $\pm$ 0.27 & 409.40  & 15.68 & 0.39 & 0.90 & x & x & x & x & x & x & x & x &  & x & x &  &  \\ 
J0353-363 & J035305.04-362308.4 & J035304.9-362307 & 0.69 $\pm$ 0.02 & 3.23 $\pm$ 0.11 & 0.69 $\pm$ 0.15 & 6.20  & 17.58 & 0.20 & 0.79 & x & x & x & x & x & x & x & x &  & x &  &  &  \\ 
J0355-187 & J035513.28-184308.5 & J035513.3-184308 & 0.61 $\pm$ 0.02 & 1.38 $\pm$ 0.09 & 0.34 $\pm$ 0.14 & 39.90  & 17.57 & 0.38 & 0.88 &  & x & x & x & x & x & x & x &  & x & x &  &  \\ 
J0356-393 & J035618.88-392141.2 & J035618.9-392141 & 0.74 $\pm$ 0.02 & 3.40 $\pm$ 0.11 & 1.43 $\pm$ 0.21 & 22.60  & 17.75 & 0.31 & 0.79 & x & x & x & x & x & x & x & x &  & x &  &  &  \\ 
J0427-185 & J042733.35-183010.3 & J042733.3-183010 & 0.97 $\pm$ 0.02 & 1.48 $\pm$ 0.10 & 0.32 $\pm$ 0.15 & 14.50  & 17.05 & 0.24 & 0.93 &  &  & x &  &  &  &  &  &  &  &  &  &  \\ 
J0443-418 & J044328.41-415156.0 & J044328.3-415156 & 0.61 $\pm$ 0.02 & 1.17 $\pm$ 0.07 & 0.23 $\pm$ 0.10 & 26.80  & 17.21 & 0.33 & 0.90 & x &  & x & x & x &  &  &  &  & x &  &  & x \\ 
J0448-165 & J044837.61-163243.2 & J044837.6-163242 & 0.83 $\pm$ 0.02 & 4.28 $\pm$ 0.17 & 2.25 $\pm$ 0.38 & 58.70  & 17.92 & 0.39 & 0.78 &  &  &  &  &  & x & x & x &  & x &  & x &  \\ 
J0452-000 & J045249.50-000152.0 & J045249.3-000153 & 0.73 $\pm$ 0.02 & 0.79 $\pm$ 0.08 & 0.63 $\pm$ 0.21 & 27.20  & 18.28 & 0.33 & 0.98 &  &  & x & x & x &  & x & x &  & x & x &  &  \\ 
J0458+089 & J045834.80+085643.4 & J045834.7+085644 & 0.67 $\pm$ 0.02 & 0.93 $\pm$ 0.09 & 0.37 $\pm$ 0.18 & 19.50  & 17.70 & 0.30 & 0.94 &  &  & x & x & x &  & x & x &  & x &  &  &  \\ 
J0504-099 & J050419.51-095632.9 & J050419.4-095632 & 0.62 $\pm$ 0.02 & 1.41 $\pm$ 0.10 & 0.57 $\pm$ 0.20 & 25.80  & 18.17 & 0.34 & 0.88 &  &  & x & x & x &  & x & x &  & x &  &  &  \\ 
J0506-545 & J050657.81-543503.7 & J050657.6-543503 & 1.30 $\pm$ 0.10 & 1.42 $\pm$ 0.06 & 0.23 $\pm$ 0.07 & 18.00  & 17.21 & 0.24 & 0.97 &  &  & x & x & x &  & x & x &  &  &  &  &  \\ 
J0509-040 & J050938.18-040045.8 & J050938.1-040046 & 0.86 $\pm$ 0.02 & 3.86 $\pm$ 0.17 & 2.17 $\pm$ 0.39 & 70.50  & 17.46 & 0.40 & 0.79 &  &  & x & x & x &  &  &  &  &  &  &  &  \\ 
J0536-337 & J053629.06-334302.5 & J053629.1-334303 & 1.98 $\pm$ 0.04 & 2.04 $\pm$ 0.10 & 0.60 $\pm$ 0.17 & 94.30  & 16.50 & 0.35 & 0.98 & x & x & x & x & x & x & x & x &  & x &  &  &  \\ 
J0543-555 & J054357.22-553207.3 & J054357.0-553204 & 2.35 $\pm$ 0.05 & 39.82 $\pm$ 0.74 & 14.61 $\pm$ 0.47 & 43.40  & 15.88 & 0.25 & 0.63 &  & x & x & x & x & x & x &  &  & x &  &  &  \\ 
J0558-386 & J055806.43-383831.5 & J055806.3-383830 & 1.47 $\pm$ 0.03 & 12.48 $\pm$ 0.24 & 3.70 $\pm$ 0.39 & 104.60  & 16.00 & 0.39 & 0.71 &  & x & x & x & x & x & x & x &  &  & x &  &  \\ 
\enddata

\vspace{0.1 cm}
\hspace{0.1 cm} {Col. (1) THC Designation\\
Cols. (2-3) \erosita and \wise names\\
Col. (4) \wise flux in the 3.4$\mu$m band \\
Col (5)  \erass flux in the 0.2 - 2.3 keV band\\
Col (6) \erass flux in the 2 - 5 keV band\\
Col (7) Radio flux from \nvss, \first, or \sumss, in order of preference. 
If the source is from \sumss, it is the flux at 843 MHz, otherwise, it is the flux at 1.4 GHz. \\
Col (8) Optical R-band magnitude taken by \textit{Gaia} \citep{GaiaCollab_2016, GaiaCollab_2023}\\ 
Cols. (9-10) Radio to infrared, and infrared to X-ray effective spectral index. \\
Cols. (11-23) A check of whether or not the source appears in the 2FHL \citep{Ackermann_2016}, 3FHL \citep{Ajello_2017}, 4FGL \citep{Abdollahi_2020}, 4LAC-DR3 \citep{Ajello2022}, 1CGH \citep{arsioli_2025}, BZCAT \citep{Massaro2009}, 3HSP \citep{Chang_2019}, 2WHSP \citep{Chang_2017}, C20 \citep{Costamante_2020}, D19 \citep{Abrusco_2019}, L25 \citep{lainez_2025}, M25 \citep{Marchesi2025}, and MAS13} \citep{Massaro_2013} catalogs. \label{table:mesh1}
\end{deluxetable*}

\section{A new list of TeV HBL candidates}
We have identified 121 TeV HBL Candidates (THC) that meet the criteria outlined in Section~\ref{sec:2}. The properties of these candidates are listed in Table~\ref{table:mesh1}. 

\begin{figure*}
    \centering
    \includegraphics[width=0.8\textwidth]{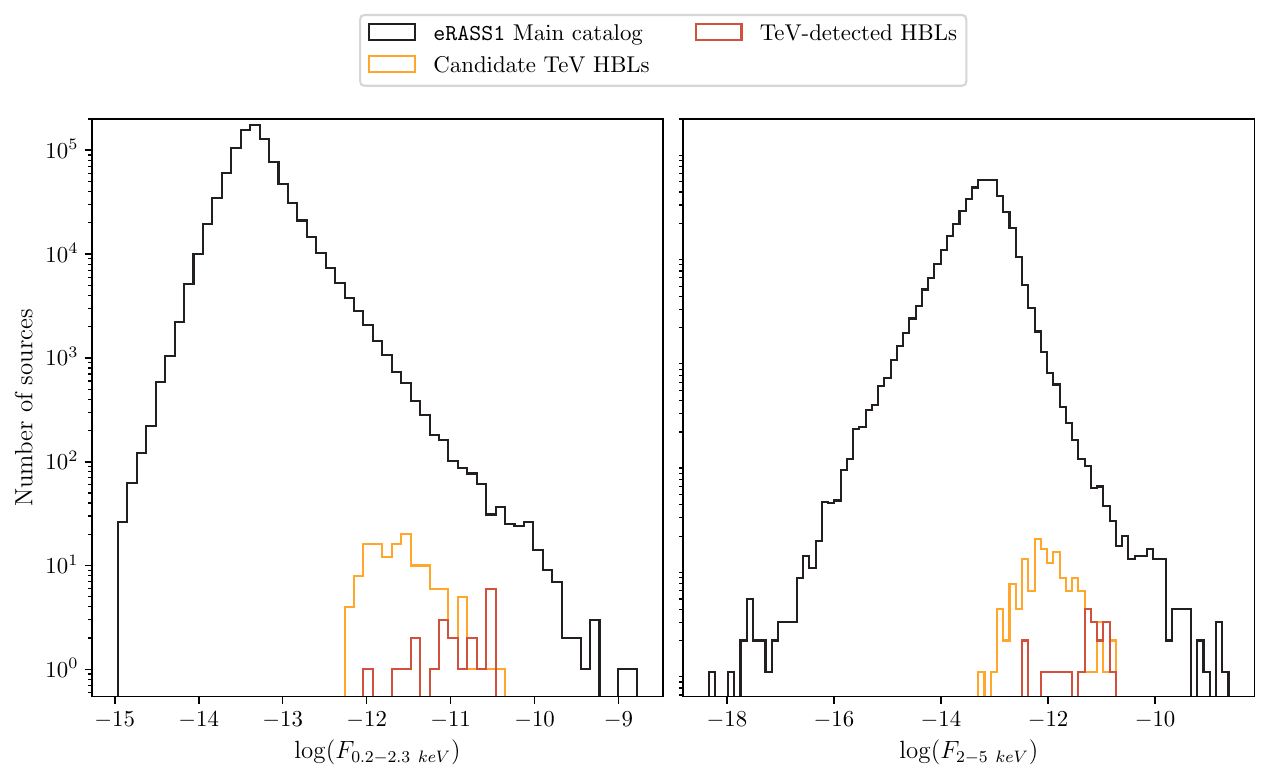}
    \caption{Distribution of X-ray fluxes for the TeV candidate blazars in different bands from the \erass\ catalog, compared to the fluxes of all sources in the eROSITA survey. }
    \label{fig:mesh:4}
\end{figure*}

X-ray and TeV fluxes show a strong correlation in TeV-emitting HBLs \citep[e.g.,][]{2005A&A...433..479K,2008ApJ...677..906F}, with the 0.2-2.3 keV flux being a good predictor for TeV detectability (Figure~\ref{fig:mesh:4}, left). By using the measured X-ray flux as an indicator of expected TeV flux, we compile a list of the most promising candidates for observation with ground-based gamma-ray observatories. These promising candidates are categorized into objects on the northern (Table~\ref{table:north}) and southern hemisphere (Table~\ref{table:south}). Most of our candidates (76\%) are located in the southern celestial hemisphere due to the larger overlap between this region and the western galactic hemisphere covered in the \erass data release. In addition, the increased \erosita exposure and subsequent increased source density around the southern ecliptic pole introduces a higher likelihood of our candidates being found on the southern hemisphere.

\begin{table}
  \centering
  \caption{Most promising TeV-peaked HBL candidates with northern declination, ranked by their 0.2-2.3\,keV flux. The last column lists sources also selected in \citet{Costamante_2020}.}
  \label{table:north}
  \begin{tabular}{ccccc}
    \hline
    Name & $F_{3.4\mu \text{m}}$ & $F_{0.2 - 2.3 \text{keV}}$ & $\alpha_{\text{IR-X}}$ & C20\\
     & mJy & $\times 10^{-12}$\,cgs & \\ \hline
J1404+040 & $1.15 \pm 0.03$ & $4.52 \pm 0.17$ & 0.81 & -\\
J0909+311 & $1.49 \pm 0.04$ & $4.42 \pm 0.23$ & 0.85 & -\\
J0837+149 & $1.91 \pm 0.04$ & $3.51 \pm 0.21$ & 0.91 & x\\
J1140+154 & $1.32 \pm 0.03$ & $3.41 \pm 0.21$ & 0.86 & x\\
J1107+150 & $0.85 \pm 0.02$ & $3.11 \pm 0.20$ & 0.82 & -\\
J1149+246 & $0.69 \pm 0.02$ & $2.95 \pm 0.19$ & 0.80 & -\\
J1324+215 & $2.78 \pm 0.06$ & $2.92 \pm 0.15$ & 0.98 & -\\
J0723+208 & $0.92 \pm 0.02$ & $2.64 \pm 0.19$ & 0.85  & -\\
J1215+075 & $2.27 \pm 0.05$ & $2.30 \pm 0.15$ & 0.98 & -\\
\hline
  \end{tabular}
\end{table}

\begin{table}
  \centering  
  \caption{Most promising TeV-peaked HBL candidates with southern declination, ranked by their 0.2-2.3\,keV flux. The last column lists sources also selected in \citet{Costamante_2020}.}
  \label{table:south}
  \begin{tabular}{ccccc}
    \hline
    Name & $F_{3.4\mu \text{m}}$ & $F_{0.2 - 2.3 \text{keV}}$ & $\alpha_{\text{IR-X}}$ & C20\\
     & mJy & $\times 10^{-12}$\,cgs &  \\ \hline
J0543-555 & $2.35 \pm 0.05$ & $39.8 \pm 0.7$ & 0.63 & - \\
J1503-156 & $0.96 \pm 0.02$ & $29.6 \pm 0.7$ & 0.55 & - \\
J0325-565 & $4.39 \pm 0.09$ & $22.9 \pm 0.5$ & 0.78 & - \\
J0156-530 & $0.85 \pm 0.02$ & $19.7 \pm 0.4$ & 0.59 & - \\
J1440-387 & $0.96 \pm 0.03$ & $13.8 \pm 0.3$ & 0.65 & - \\
J1548-228 & $3.21 \pm 0.07$ & $13.0 \pm 0.3$ & 0.81 & - \\
J0209-524 & $2.54 \pm 0.05$ & $12.9 \pm 0.3$ & 0.78 & - \\
J0558-386 & $1.47 \pm 0.03$ & $12.5 \pm 0.2$ & 0.71 & - \\
J0123-231 & $0.73 \pm 0.02$ & $11.9 \pm 0.3$ & 0.63 & - \\
J0244-583 & $1.67 \pm 0.04$ & $10.6 \pm 0.2$ & 0.24 & x \\
    \hline
  \end{tabular}
\end{table}

The probability of spurious spatial coincidences between an infrared-selected HBL-like object and an X-ray source in the \erass catalog is negligible, leading to fewer than one expected spurious candidate in our sample. 
Using 100,000 random sky points placed within the \texttt{eFEDS} field \citep{eFEDS}, which covers approximately $\sim140\,\text{deg}^2$, the random spatial correlation rate is 0.117\%. Given that the \texttt{eFEDS} exposure is approximately 10 times that of the \erass survey, we anticipate an even lower probability of random spatial association in the \erass data set.

As part of our selection algorithm, we rejected objects with infrared properties similar to those of TeV-emitting HBLs that have an \erosita counterpart but were labeled as an extended X-ray source. The 31 sources excluded from our list for this reason are listed in Table~\ref{table:extended}. These could still be objects of interest, as some of them could be associated with BL~Lac-type blazars embedded in the core of a cluster of galaxies (similar to NGC~1275 and IC~310, located in the Perseus cluster), with the intracluster emission causing the X-ray counterpart to be extended in the \erosita data. In addition, we discarded 18 sources because a radio counterpart could no be identified in the \texttt{NVSS}, \texttt{FIRST}, \texttt{SUMSS}, or other \texttt{RADIO} catalogs. These objects, listed in Table~\ref{table:noRadio}, could merit dedicated radio follow-up observations to determine whether a weak radio counterpart can be identified.

\begin{table*}
  \centering  
  \caption{New blazar candidates, ordered by Right Ascension. The abbreviations for the \texttt{Simbad} counterparts are: Active Galactic Nucleus (AGN), Brightest Galaxy in a Cluster (BiC), Galaxy (G), Radio Galaxy (rG), Seyfert 1 Galaxy (Sy1), and X-ray source (X).}
  \label{table:new_blazar_candidates}
  \begin{tabular}{cccccc}
    \hline
    Name & $F_{3.4\mu \text{m}}$ & $F_{0.2 - 2.3 \text{keV}}$ & $\alpha_{\text{IR-X}}$ & $R_\mathrm{mag}$ & \texttt{Simbad} counterpart\\
     & mJy & $\times 10^{-12}$\,cgs &  &  &object type\\ \hline
J0117-549  &  $0.96\pm0.02$  &  $1.18\pm0.09$ &  0.96 & 17.36 &  BiC  \\
J0356-393  &  $0.74\pm0.02$  &  $3.40\pm0.11$ &  0.79 &  17.75& G  \\
J0803+193  &  $0.89\pm0.02$  &  $1.06\pm0.12$ &  0.96 &  17.03 & rG  \\
J0824+016  &  $1.22\pm0.03$  &  $1.49\pm0.14$ &  0.96 &  17.43 & X  \\
J0914+088  &  $0.85\pm0.04$  &  $1.64\pm0.14$ &  0.90 & 17.87 & AGN, Sy1  \\
J0914+235  &  $0.60\pm0.02$  &  $0.84\pm0.10$ &  0.94 &  18.66 & -  \\
J1324+215  &  $2.78\pm0.06$  &  $2.92\pm0.15$ &  0.98 & 16.37 & Sy1  \\
J1344-451  &  $2.95\pm0.06$  &  $3.04\pm0.13$ &  0.98 & 16.97 & G  \\
    \hline
  \end{tabular}
\end{table*}

\section{Discussion and conclusions}\label{sec:conclusions}
Our search for TeV-emitting HBL candidates based on the \wise and \erosita surveys has led to the identification of 121 promising objects. Among these, 53 (43\%) are identified as BL~Lacs in the ROMA-BZCAT catalog \citep{Massaro2009,2015ApSS.357...75M}. 
Our search finds a large overlap with the \texttt{3HSP} catalog \citep{Chang_2019}, which identified 2013 HBL blazar candidates based on radio-to-X-ray effective spectral indices and a radio flux cut. Specifically, 108 of the 121candidates (89\%) are also featured in the \texttt{3HSP} catalog. The \texttt{3HSP} catalog selects 2013 sources with $\nu_{\text{peak}}\gtrsim 4$\,eV 
and goes an order of magnitude deeper in X-ray flux ($4\times 10^{-14}$\,\fluxcgs) than the weakest X-ray source in our final candidate list ($6\times 10^{-13}$\,\fluxcgs). Half of our candidates not present in the \texttt{3HSP} catalog were not detected by \rosat. In total, 23 of our HBL candidates (19\%) do not have a \rosat counterpart \citep[\texttt{2RXS} catalog,][]{2016A&A...588A.103B}, showing the power that the depth of the \erass survey has to reveal new TeV HBL candidates that may have eluded detection in previous searches.

\begin{figure*}[t]
    \centering
    \includegraphics[width=0.8\textwidth]{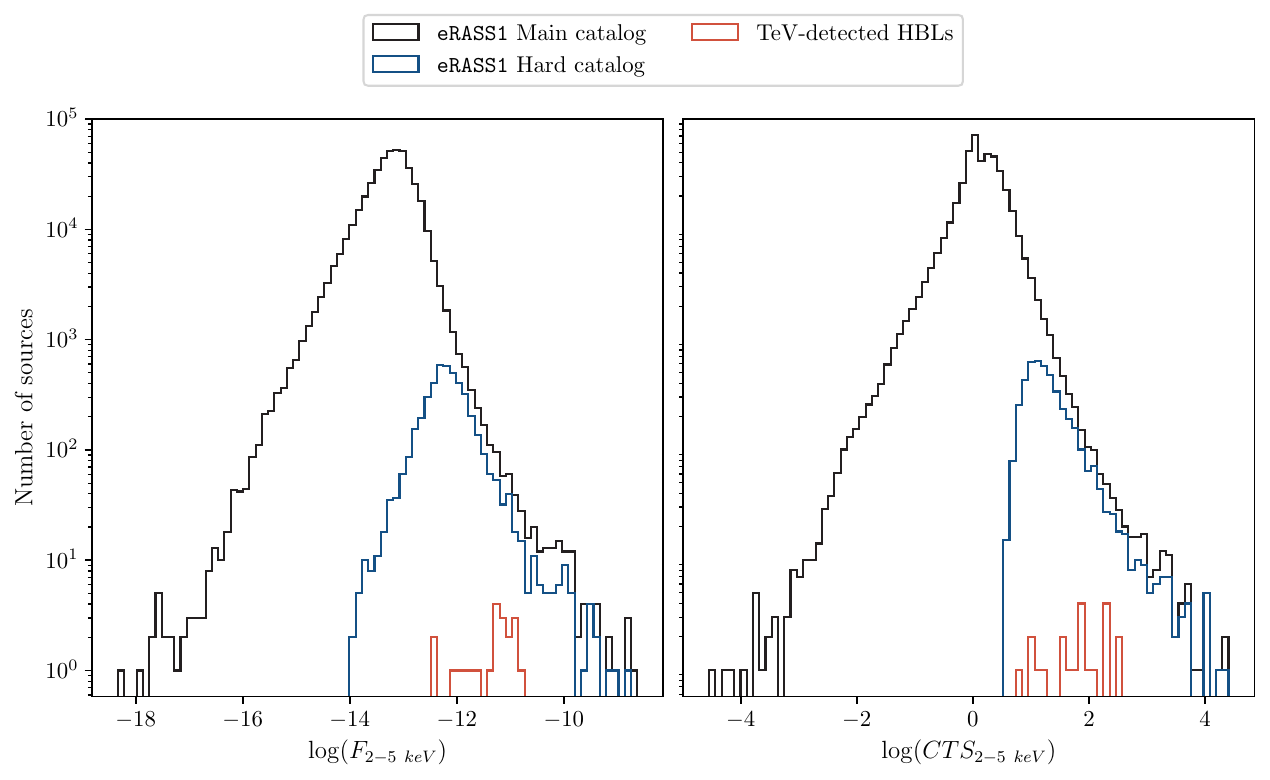}
    \caption{The flux and counts distribution in the 2.0-5.0 keV band for the main \erass catalog (black) and hard \erass catalog (blue) compared with the flux and counts for TeV-detected HBLs (red).}
    \label{catalog_comparisons}
\end{figure*}

A second notable advantage of \erosita over \rosat is its extended energy range, covering higher X-ray energies. \erosita is sensitive in the 0.2–8\,keV band, whereas \rosat was limited to the 0.1–2.4\,keV range. The initial data release from \erosita includes two catalogs: a main catalog of sources detected between 0.2 and 2.3\,keV, and a hard source catalog for sources detected between 2.3 and 5\,keV \citep{Merloni_2024}.
While all 21 TeV-detected HBLs in the western galactic hemisphere were significantly detected in the main \erass catalog, two were not found in the \erosita hard catalog. Figure~\ref{catalog_comparisons} illustrates that some TeV HBLs appear at hard X-ray fluxes below the threshold at which the \erass hard catalog can be considered complete. 
Conversely, the \erass main source catalog appears to be complete down to the flux levels of the weakest TeV-detected HBLs.
For this reason, we used the main \erass catalog, rather than the hard catalog, to characterize the X-ray emission of our source candidates.

Our list of HBL candidates includes 11 objects that have not been previously identified as blazars in the \texttt{4LAC}, \texttt{BZCAT} and \texttt{3HSP} catalogs.  
However, three of these are included in other blazar candidate catalogs (\texttt{BROS}, \citet{2020ApJ...901....3I}; \texttt{KDEBLLACS} \& \texttt{WIBRaLS2}, \citet{Abrusco_2019}) and are also listed in the eROSITA blazar catalog by H\"ammerich et al. (in prep).
As a result, the remaining 8 sources are considered new blazar candidates. Table~\ref{table:new_blazar_candidates} summarizes the infrared and X-ray properties on these new blazar candidates, together with the source type of \texttt{Simbad} counterparts within 10'' of the \texttt{WISEA} source. 
One source (THC~J0914+235) does not have a \texttt{Simbad} 
within our search radius. 
THC~J0825+016 is spatially coincident with an X-ray source for which no other multi-wavelength identification exists.
For the remaining sources, we find extragalactic counterparts, for some of which AGN activity has been reported. One source (THC\,J0914+088) has two potential counterparts, that is, an AGN and a Seyfert 1 galaxy, which are both found to have a redshift of $z=0.14$ and have been proposed to be interacting with one another \citep{2021AJ....162..276Z}.

Our TeV-peaked HBL candidates are also found prominently in catalogs that use the high-energy response of \lat to find GeV-bright objects: 31 are listed in a search for $>50$\,GeV sources using 6.6 years of LAT data \citep[\texttt{2FHL}, ][]{Ackermann_2016}, 65 in the \texttt{3FHL} \citep[$>10$\,GeV, 7 years, ][]{Ajello_2017}, and 103 in the \texttt{1CGH} catalog \citep[$>10$\,GeV, 7 years, ][]{arsioli_2025}.

\begin{figure}[t]
    \centering
    \includegraphics[width=0.47\textwidth]{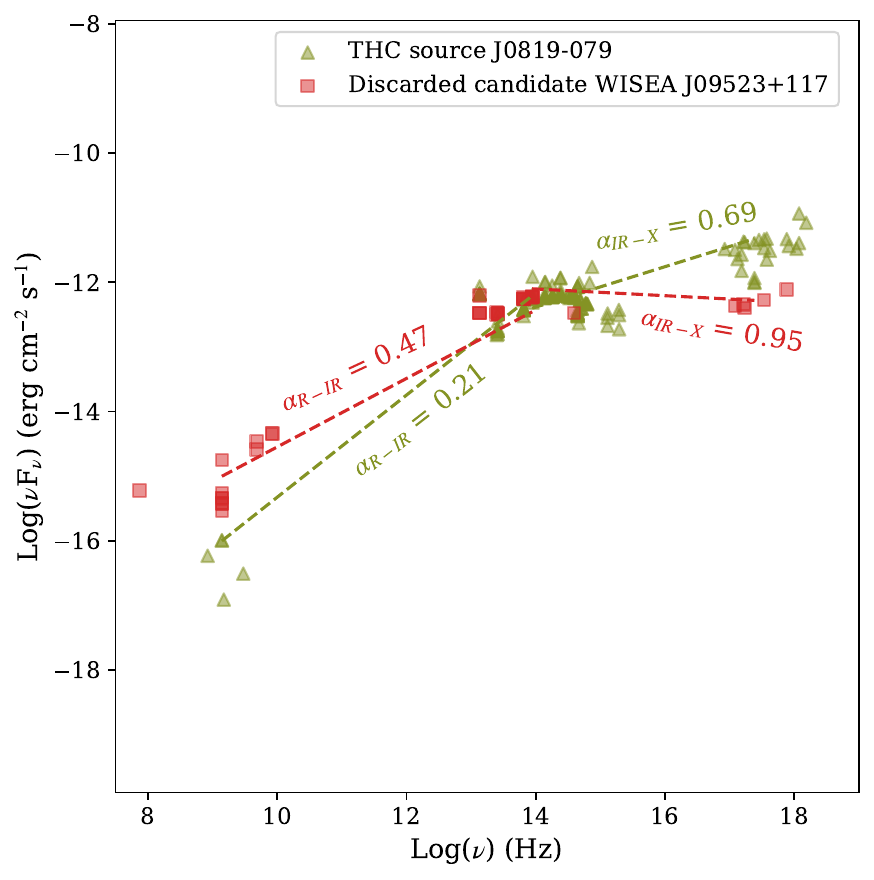}
    \caption{Example of the synchrotron SED of selected source THC J0819-079 (green triangles) compared to potential candidate WISEA J095233.24+114520.4 that is discarded by the $\alpha_{\text{R-IR}}$ cut (red squares). Quasar-like sources can make it through our selection criteria based on infrared flux and spectrum and display an $\alpha_{\text{IR-X}}$ similar to that of TeV HBLs even though their synchrotron peak is probably in the infrared band and their X-ray flux, showing a hard very hard spectrum, samples the rising edge of their high-energy spectral component. These quasar-like objects will tend to display larger jet powers than BL~Lacs, and their excess radio flux can be used to discriminate them from BL~Lacs using the $\alpha_{\text{R-IR}}$ criterion (See Section~\ref{sec:2}).}
    \label{fig:sed_comp}
\end{figure}

A recent, dedicated search for TeV-peaked BL Lac candidates with a gamma-ray peak $>1$\,TeV utilizes X-ray flux and the effective spectral index between the X-ray and gamma-ray bands to identify 47 candidates \citep{Costamante_2020}. This search is particularly effective in finding candidates with low radio flux.
Among the 16 sources identified by \citet{Costamante_2020} located in the western galactic hemisphere, 6 are also found in our search. On the other hand, our algorithm did not select 10 sources listed in \citet{Costamante_2020}: six fail the $M_{3.4 \mu \text{m}}$ threshold, and three do not fit within the infrared color-color selection criteria. The remaining source narrowly misses our $\alpha_{\text{IR-X}}$ cut. The candidates with lower radio flux in \citet{Costamante_2020} are likely to exhibit lower infrared flux as well, making our searches complementary. 

In our study, we have introduced the radio-IR-X spectral slopes to separate flat spectrum radio quasars from BL~Lacs (Figure~\ref{fig:sourcetypes}). As seen in Figure~\ref{fig:sed_comp}, quasar-like sources can meet our selection criteria that rely on infrared flux and spectral features, and display an $\alpha_{\text{IR-X}}$ similar to that found in TeV HBLs. This happens even though their synchrotron peak is likely in the infrared range, while their X-ray flux, with a particularly hard spectrum, samples the rising phase of their high-energy spectral emission. These quasar-like sources typically have more powerful jets compared to BL Lacs, and their additional radio flux distinguishes them from BL Lacs using the $\alpha_{\text{R-IR}}$ criterion. In addition, extreme HBLs with a high-energy spectral component peaking beyond 1\,TeV do appear to have higher $\alpha_{\text{IR-X}}$ values than regular TeV-detected HBLs (Figure~\ref{fig:sourcetypes}), yielding yet another metric that can be used to select extreme objects. 

HBLs are brightest in the X-ray band due to the location of their synchrotron peak. Our study takes advantage of the recently released \erass catalog, which currently constitutes the deepest survey where HBLs can be found. The \lat allsky survey in the gamma-ray band is similarly deep. 
However, with LAT sensitivity peaking at a few GeV, high LAT fluxes select sources with soft GeV spectra rather than the hard GeV spectra expected for HBLs. As noted by \citet{Costamante_2020}, the most extreme blazars might not appear in radio loud AGN catalogs due to their low jet power and consequently weak radio emission compared to the bright optical flux from the host galaxy, reducing their radio loudness parameter.
Current and future image atmospheric Cherenkov telescopes continue to be the most efficient means to find and characterize TeV-emitting BL Lacs. On the other hand, gamma-ray observatories based on particle-sampling arrays have the advantage of a large field of view and efficient duty cycle, carrying out effective surveys of the TeV sky. However, their limited sensitivity below 1 TeV reduces their effectiveness for objects with redshifts $z\gtrsim 0.1$ due to the absorption of TeV gamma rays by the extragalactic background light.
LHAASO has detected gamma-ray emission from four blazars \citep{2024ApJS..271...25C}, while HAWC has significant detections of Mrk 241 and Mrk 501, along with evidence for emission from three other blazars \citep{2021ApJ...907...67A}. The results of our study, listed in Table~\ref{table:mesh1}, offer promising candidates for pointed searches for new TeV-emitting blazars, which can be conducted by HESS, VERITAS, MAGIC and LST-1, as well as by CTAO once it begins its operations.

\facilities{\textit{eROSITA}, WISE.}

\software{astropy \citep{2013A&A...558A..33A,2018AJ....156..123A,2022ApJ...935..167A},  
HEASoft \citep{2014ascl.soft08004N}}.



\begin{acknowledgments}
CM and AG acknowledge financial support from the McDonnell Center for the Space Sciences at Washington University in Saint Louis that made this study possible. The authors want to thank the anonymous referee, as well as Raffaele D'Abrusco, Jamie Holder, Stefano Marchesi, Steven H\"{a}mmerich, and Felicia McBride for valuable feedback on the first draft of this paper.
\end{acknowledgments}



\bibliography{references}
\bibliographystyle{aasjournal}
\appendix

In this Appendix, we include three tables as supplementary information. Table~\ref{table:TeV} lists the properties of TeV-detected HBLs. Table ~\ref{table:extended} lists infrared sources of interest that were spatially associated with an extended X-ray counterpart in the \erosita survey. Table~\ref{table:noRadio} lists infrared sources of interest with an \erosita counterpart for which we could not find a radio counterpart in the \texttt{NVSS}, \texttt{FIRST}, \texttt{SUMSS}, or other \texttt{RADIO} catalogs.

\begin{deluxetable*}{ccccccc} 

\tabletypesize{\scriptsize}  
\tablenum{A1}
\label{table:TeV}

\tablecaption{Radio, infrared and X-ray properties of TeV-detected HBLs, ordered by RA.}

\tablehead{\colhead{Name} & \colhead{$F_{3.4\mu\text{m}}$}  & \colhead{$F_{0.2 - 2.3 keV}$} & \colhead{$F_{2 - 5 keV}$}  & \colhead{Radio Flux}& \colhead{$\alpha_{\text{R-IR}}$} & \colhead{$\alpha_{\text{IR-X}}$} \\
\colhead{} & \colhead{mJy} & \colhead{$\times 10^{-12}$ cgs} & \colhead{$\times 10^{-12}$ cgs} & \colhead{mJy} & \colhead{} & \colhead{} 
}
\startdata
1ES 1028+511 & 1.40 $\pm$ 0.03 & -- & -- & 37.5 & 0.30 & -- \\ 
RBS 1366 & 1.54 $\pm$ 0.05 & -- & -- & 39.5 & 0.29 & -- \\ 
MRC 0910-208 & 3.96 $\pm$ 0.08 & 11.36 $\pm$ 0.33 & 5.94 $\pm$ 0.71 & 328.4 & 0.40 & 0.85 \\ 
1RXS J195815.6-301119 & 2.87 $\pm$ 0.06 & -- & -- & 127.5 & 0.34 & -- \\ 
1RXS J081201.8+023735 & 1.53 $\pm$ 0.04 & 3.18 $\pm$ 0.20 & 0.89 $\pm$ 0.31 & 77 & 0.40 & 0.89 \\ 
TXS 1515-273 & 5.17 $\pm$ 0.11 & 1.16 $\pm$ 0.10 & 0.40 $\pm$ 0.18 & 564.2 & 0.42 & 1.17 \\ 
TXS 0210+515 & 7.46 $\pm$ 0.16 & -- & -- & 294.3 & 0.33 & -- \\ 
1ES 2322-409 & 7.13 $\pm$ 0.15 & 8.92 $\pm$ 0.29 & 2.53 $\pm$ 0.47 & 49 & 0.18 & 0.96 \\ 
PGC 2402248 & 4.13 $\pm$ 0.09 & -- & -- & 23.5 & 0.16 & -- \\ 
1RXS J023832.6-311658 & 2.99 $\pm$ 0.06 & 30.28 $\pm$ 1.10 & 12.53 $\pm$ 0.75 & 71 & 0.29 & 0.69 \\ 
1ES 2037+521 & 6.12 $\pm$ 0.12 & -- & -- & 40.7 & 0.17 & -- \\ 
RGB J2243+203 & 5.29 $\pm$ 0.12 & -- & -- & 108.5 & 0.27 & -- \\ 
RX J1136.5+6737 & 2.05 $\pm$ 0.04 & -- & -- & 45.3 & 0.28 & -- \\ 
RBS 0723 & 1.20 $\pm$ 0.03 & 14.84 $\pm$ 0.43 & 6.79 $\pm$ 0.88 & 32.8 & 0.30 & 0.67 \\ 
MS 1221.8+2452 & 1.40 $\pm$ 0.03 & 3.56 $\pm$ 0.18 & 1.21 $\pm$ 0.33 & 24.5 & 0.26 & 0.87 \\ 
H 1722+119 & 6.96 $\pm$ 0.14 & -- & -- & 120.4 & 0.26 & -- \\ 
KUV 00311-1938 & 3.74 $\pm$ 0.08 & -- & -- & 18.5 & 0.14 & -- \\ 
PKS 0301-243 & 10.23 $\pm$ 0.21 & 2.09 $\pm$ 0.10 & 0.38 $\pm$ 0.13 & 700.2 & 0.38 & 1.18 \\ 
RGB J0136+391 & 5.63 $\pm$ 0.12 & -- & -- & 60 & 0.21 & -- \\ 
PKS 1440-389 & 7.26 $\pm$ 0.15 & 4.26 $\pm$ 0.15 & 1.65 $\pm$ 0.28 & 109.8 & 0.25 & 1.05 \\ 
1ES 1727+502 & 4.85 $\pm$ 0.10 & -- & -- & 200.7 & 0.34 & -- \\ 
1ES 0033+595 & 4.80 $\pm$ 0.10 & -- & -- & 147.3 & 0.31 & -- \\ 
1ES 0647+250 & 4.48 $\pm$ 0.10 & 22.33 $\pm$ 0.54 & 10.00 $\pm$ 1.08 & 96.2 & 0.28 & 0.78 \\ 
1ES 1741+196 & 6.86 $\pm$ 0.15 & -- & -- & 301.2 & 0.34 & -- \\ 
1ES 1215+303 & 13.16 $\pm$ 0.27 & 7.64 $\pm$ 0.30 & 1.30 $\pm$ 0.36 & 571.6 & 0.34 & 1.05 \\ 
1RXS J101015.9-311909 & 2.55 $\pm$ 0.06 & 9.62 $\pm$ 0.31 & 5.06 $\pm$ 0.67 & 73.5 & 0.30 & 0.82 \\ 
1ES 1312-423 & 1.67 $\pm$ 0.04 & 9.29 $\pm$ 0.22 & 5.33 $\pm$ 0.49 & 13.9 & 0.18 & 0.77 \\ 
SHBL J001355.9-185406 & 3.02 $\pm$ 0.09 & -- & -- & 29.2 & 0.21 & -- \\ 
HESS J1943+213 & 1.85 $\pm$ 0.20 & -- & -- & 102.6 & 0.36 & -- \\ 
B3 2247+381 & 4.69 $\pm$ 0.10 & -- & -- & 103.4 & 0.28 & -- \\ 
1ES 1440+122 & 2.00 $\pm$ 0.06 & -- & -- & 68.8 & 0.32 & -- \\ 
RX J0648.7+1516 & 2.52 $\pm$ 0.06 & 6.61 $\pm$ 0.29 & 4.43 $\pm$ 0.72 & 64.2 & 0.29 & 0.86 \\ 
PKS 0447-439 & 18.96 $\pm$ 0.38 & 32.64 $\pm$ 0.36 & 6.23 $\pm$ 0.47 & 382.1 & 0.26 & 0.92 \\ 
1ES 0414+009 & 2.54 $\pm$ 0.06 & 17.75 $\pm$ 0.47 & 13.49 $\pm$ 2.02 & 74 & 0.34 & 0.74 \\ 
1ES 0502+675 & 2.48 $\pm$ 0.05 & -- & -- & 25.4 & 0.21 & -- \\ 
RBS 0413 & 1.32 $\pm$ 0.03 & -- & -- & 20.9 & 0.25 & -- \\ 
PKS 1424+240 & 28.22 $\pm$ 0.60 & -- & -- & 429.7 & 0.25 & -- \\ 
RGB J0710+591 & 2.79 $\pm$ 0.06 & -- & -- & 63 & 0.28 & -- \\ 
1ES 0806+524 & 7.23 $\pm$ 0.15 & -- & -- & 182.4 & 0.29 & -- \\ 
RGB J0152+017 & 5.49 $\pm$ 0.12 & -- & -- & 61.4 & 0.22 & -- \\ 
1ES 1011+496 & 7.88 $\pm$ 0.17 & -- & -- & 377.7 & 0.35 & -- \\ 
1ES 0347-121 & 0.68 $\pm$ 0.02 & 29.81 $\pm$ 0.74 & 15.27 $\pm$ 0.92 & 24 & 0.32 & 0.51 \\ 
PKS 0548-322 & 4.29 $\pm$ 0.09 & 19.19 $\pm$ 0.36 & 8.54 $\pm$ 0.65 & 343.7 & 0.40 & 0.79 \\ 
1ES 0229+200 & 2.55 $\pm$ 0.06 & -- & -- & 82.4 & 0.31 & -- \\ 
Markarian 180 & 10.95 $\pm$ 0.21 & -- & -- & 327.1 & 0.31 & -- \\ 
1ES 1218+304 & 4.62 $\pm$ 0.09 & 26.95 $\pm$ 0.55 & 6.82 $\pm$ 0.82 & 71 & 0.25 & 0.76 \\ 
H 2356-309 & 2.08 $\pm$ 0.05 & -- & -- & 62.1 & 0.31 & -- \\ 
1ES 1101-232 & 2.22 $\pm$ 0.05 & 29.88 $\pm$ 0.55 & 12.93 $\pm$ 1.01 & 66 & 0.35 & 0.66 \\ 
PG 1553+113 & 21.74 $\pm$ 0.46 & -- & -- & 312 & 0.24 & -- \\ 
PKS 2005-489 & 26.51 $\pm$ 0.56 & 27.61 $\pm$ 0.56 & 6.41 $\pm$ 0.80 & 1299 & 0.34 & 0.98 \\ 
H 1426+428 & 2.54 $\pm$ 0.06 & -- & -- & 58 & 0.28 & -- \\ 
1ES 1959+650 & 14.58 $\pm$ 0.31 & -- & -- & 249.6 & 0.26 & -- \\ 
PKS 2155-304 & 64.88 $\pm$ 1.37 & -- & -- & 489.3 & 0.18 & -- \\ 
1ES 2344+514 & 12.22 $\pm$ 0.25 & -- & -- & 250.4 & 0.27 & -- \\ 
Markarian 501 & 36.35 $\pm$ 0.74 & -- & -- & 1558 & 0.34 & -- \\ 
Markarian 421 & 53.28 $\pm$ 1.08 & -- & -- & 767.4 & 0.24 & -- \\ 
\enddata

\vspace{0.1 cm} \hspace{0.1 cm} {Col. (1) Name.\\
Col. (2) \wise flux in the $3.4 \mu \text{m}$ band.\\
Col. (3) \erosita flux in the $0.2 - 2.3 \text{keV}$ band. \\
Col. (4) \erosita flux in the $2 - 5 \text{keV}$ band. \\
Col. (5) The radio flux from \sumss, \nvss, or \first. If the source is found in all 3 catalogs, the \nvss flux is listed, if the source is found in only \sumss and \first, the \first flux is listed. If the source is from \sumss, it is the flux at 843 MHz, otherwise, it is the flux at 1.4 GHz. \\
Col. (6)  The effective spectral index radio to infrared. \\
Col (7) The effective spectral index infrared to X-ray as measured with \erosita (not given for sources in the Eastern hemisphere). 
}

\end{deluxetable*}

\begin{table}
\tablenum{A2}
\label{table:extended}
  \centering
  \caption{List of \wise-\erosita sources that are discarded as potential TeV candidates due to the \erosita counterpart being flagged as an extended X-ray source in the \erass catalog.}
  \begin{tabular}{ll}
    \hline
    \wisea & \erass  \\ \hline
J021616.62-481625.9 & J021616.7-481624 \\ 
J024235.93-213225.6 & J024235.9-213225 \\ 
J031757.67-441417.0 & J031757.8-441417 \\ 
J040422.61-651013.0 & J040422.7-651016 \\ 
J043902.25+052043.6 & J043902.2+052039 \\ 
J102557.93+124103.2 & J102557.9+124108 \\ 
J113250.86+142738.6 & J113251.1+142739 \\ 
J114124.20-121638.5 & J114124.0-121639 \\ 
J152412.92-315422.6 & J152412.8-315421 \\ 
J153014.29-423151.7 & J153014.1-423147 \\ 
J180228.51-523645.6 & J180228.7-523646 \\ 
J013803.72-215531.2 & J013803.9-215530 \\ 
J023727.86-263028.7 & J023727.6-263031 \\ 
J051031.94-183844.0 & J051031.7-183842 \\ 
J051636.92+062648.4 & J051636.9+062648 \\ 
J071023.81+224002.4 & J071023.8+224004 \\ 
J073720.91+351741.8 & J073721.1+351739 \\ 
J080056.86+360324.9 & J080056.8+360324 \\ 
J081545.88+100411.4 & J081545.8+100407 \\ 
J083759.97-425926.7 & J083759.7-425925 \\ 
J084255.97+292727.0 & J084255.8+292722 \\ 
J100638.78+255440.8 & J100639.0+255443 \\ 
J104432.88-070406.3 & J104432.9-070406 \\ 
J105036.50-023616.3 & J105036.3-023618 \\ 
J110331.50-083511.1 & J110331.3-083513 \\ 
J111111.17-300922.1 & J111111.2-300924 \\ 
J111450.28-121350.5 & J111450.1-121353 \\ 
J113542.70-202105.8 & J113542.4-202109 \\ 
J121259.25+272704.1 & J121259.1+272708 \\ 
J150722.47-432324.6 & J150722.4-432328 \\ 
J183659.31-664907.7 & J183658.7-664905 \\
\hline
  \end{tabular}
\end{table}

\begin{table}
\tablenum{A3}
\label{table:noRadio}
  \centering
  \caption{List of \wise-\erosita sources that are discarded as potential TeV candidates because a radio counterpart within $5''$ is not found.}
  \begin{tabular}{ll}
    \hline
    \wisea & \erass  \\ \hline
J015721.56-161420.9 & J015721.4-161420 \\ 
J010710.08-321655.0 & J010709.8-321654 \\ 
J014100.48-675327.3 & J014100.5-675326 \\ 
J030246.33-192425.3 & J030246.3-192426 \\ 
J030321.79-633820.6 & J030321.6-633819 \\ 
J043509.66-752743.8 & J043509.7-752743 \\ 
J063015.06-201236.4 & J063014.9-201234 \\ 
J064047.66-242313.7 & J064047.6-242313 \\ 
J070819.92-412333.7 & J070819.6-412333 \\ 
J074457.94-525713.6 & J074457.9-525714 \\ 
J081749.19+025059.3 & J081749.1+025058 \\ 
J105037.90+170208.3 & J105038.1+170209 \\ 
J103946.98-050658.5 & J103947.2-050657 \\ 
J112019.17-101843.5 & J112019.3-101842 \\ 
J113826.75+032206.4 & J113826.8+032205 \\ 
J122337.01-303250.1 & J122337.1-303249 \\ 
J123137.69-493423.9 & J123137.7-493424 \\ 
J124557.42-125119.6 & J124557.4-125118 \\ 
J135120.21+033716.5 & J135120.1+033716 \\ 
J140907.20-451715.9 & J140907.2-451716 \\ 
J143342.76-730438.1 & J143342.7-730435 \\ 
J144909.23-090815.4 & J144909.3-090815 \\ 
J150654.76-215925.5 & J150654.7-215925 \\ 
J213237.80-623750.9 & J213237.9-623750 \\ 
J230647.91-385603.3 & J230648.0-385603 \\ 
J020608.36-774343.7 & J020608.1-774343 \\ 
J033334.62-425552.3 & J033334.6-425552 \\ 
J051541.41+010440.3 & J051541.4+010440 \\ 
J071104.52-493023.2 & J071104.2-493023 \\ 
J074722.17+090548.0 & J074722.2+090546 \\ 
J084853.09+282411.8 & J084852.9+282415 \\ 
J092655.25+082249.2 & J092655.3+082247 \\ 
J101121.18-021239.7 & J101121.0-021241 \\ 
J122534.95+073025.7 & J122534.8+073028 \\ 
J132838.69+120952.5 & J132838.6+120953 \\ 
J202943.64-552247.9 & J202943.8-552245 \\ 

\hline
  \end{tabular}
\end{table}
\end{document}